\documentclass[12pt]{iopart}

\usepackage{iopams}

\expandafter\let\csname equation*\endcsname=\relax
\expandafter\let\csname endequation*\endcsname=\relax
\usepackage{amsmath}
\usepackage{mathtools}
\usepackage{tensor}
\usepackage{stmaryrd} 

\usepackage{pdflscape}
\usepackage{lscape}
\usepackage{array}
\usepackage[usenames,dvipsnames]{xcolor}
\IfFileExists{aastexv6.sty}{\input{aastexv6.sty}}{\input{/Users/quentinvigneron/Documents/Travail/Research/tex_/aastexv6.sty}}
\IfFileExists{Short_cut.sty}{\input{Short_cut.sty}}{\input{/Users/quentinvigneron/Documents/Travail/Research/tex_/Short_cut.sty}}

 \usepackage[numbers,sort&compress]{natbib}
\usepackage{hyperref}

\usepackage[framemethod=TikZ]{mdframed} 
\mdfsetup{%
   middlelinecolor=black, 
   middlelinewidth=1pt,
   backgroundcolor=black!15, 
   roundcorner=15pt}

\begin{document}

\title[On non-Euclidean Newtonian theories]{On non-Euclidean Newtonian theories and their cosmological backreaction}

\author{Quentin Vigneron$^{1}$}

\address{$^1$ Institute of Astronomy, Faculty of Physics, Astronomy and Informatics, Nicolaus Copernicus University, Grudziadzka 5, 87-100 Toru\'n, Poland}
\ead{quvigneron@gmail.com, qvignero@astro.uni.torun.pl}

\vspace{10pt}
\begin{indented}
\item[]\today
\end{indented}

\begin{abstract}

Constructing an extension of Newton's theory which is defined on a non-Euclidean topology (in the sense of Thurston's decomposition), called \textit{a non-Euclidean Newtonian theory}, corresponding to the zeroth order of a non-relativistic limit of general relativity is an important step in the study of the backreaction problem in cosmology and might be a powerful tool to study the influence of global topology on structure formation. After giving a precise mathematical definition of such a theory, based on the concept of Galilean manifolds, we propose two such extensions, for spherical or hyperbolic topologies, using a minimal modification of the Newton-Cartan equations. However as for now we do not seek to justify this modification from general relativity. The first proposition features a non-zero cosmological backreaction, but the presence of gravitomagnetism and the impossibility of performing exact $N$-body calculations make this theory difficult to be interpreted as a Newtonian-like theory. The second proposition features no backreaction, exact $N$-body calculation is possible and no gravitomagnetism appears. In absence of a justification from general relativity, we argue that this non-Euclidean Newtonian theory should be the one to be considered, and could be used to study the influence of topology on structure formation via $N$-body simulations. For this purpose we give the mass point gravitational field in $\mS^3$.

\end{abstract}

%
%
%
%
%

\section{Introduction}

The cosmological backreaction corresponds to the effect of small scale inhomogeneities on the global expansion of the Universe. The main question surrounding this phenomenon is whether or not it can produce observable effects on the scale factor evolution, in particular an acceleration of its evolution \cite{2008_Buchert} or an effect which might explain the discrepancy in the measure of the Hubble constant \cite{2020_Heinesen_et_al}. One of the main result obtained about backreaction is the {\BET} \cite{1997_Buchert_et_al, 2021_Vigneron}, stating that this phenomenon is exactly zero when calculated using Newton's theory of gravitation
, which may not be the case in general relativity. This seems to indicate that in a universe whose structure formation is well described by Newton's theory, backreaction should be negligible.

As pointed out by \cite{2022_Vigneron}, this statement is valid only if that universe has a Euclidean topology (we define this term in section~\ref{sec::topo}), i.e. the topology on which Newton's theory is defined, implying that the {\BET} needs to be generalised for non-Euclidean topologies. This generalisation requires an extension of Newton's theory to such topologies, called \textit{a non-Euclidean Newtonian theory} (hereafter NEN theory). As defined in \cite{2022_Vigneron}, the equations in this theory should reduce to the classical Newton's equations on small scales (locally Newtonian theory) but be defined on a non-Euclidean topology.

We stress that the purpose of this modification of Newton's theory is not to take into account relativistic effects of spatial curvature that would come from general relativity (as in e.g. \citep{2014_Abramowicz_et_al}), but effects of the global topology. This latter aspect of cosmology can indeed not only be described by a relativistic theory, i.e. invariant under Lorentzian transformations, but also by a Newtonian theory, i.e. invariant under (local) Galilean transformations.

Another motivation than the study of backreaction for the construction of a NEN theory is the study of structure formation. As for now most of the simulations of this phenomenon are done using Newton's theory, mostly for its simplicity with respect to general relativity. These simulations are thus limited to model universes with a (closed) Euclidean topology, which is generally the 3-torus $\mathbb{T}^3$. A NEN theory would be a powerful tool for cosmology as it would keep the simplicity of Newton's theory, but would allow for the study of structure formation in non-Euclidean topologies, in a sense allowing for other types of periodic boundary conditions than the usual periodic cube. In particular, such a theory would assess whether or not observable effects of topology could be visible on the dynamics and morphology of structures, which as for now, has never been studied to our knowledge. This might provide us with a new method to detect the precise topology of our Universe, other than the existing cosmic crystallography \cite{1996_Le_La_Lu} and ``circle in the sky'' (e.g. \cite{1999_Luminet_et_al} and \cite{2003_Luminet_et_al}) methods.

To our knowledge, only one NEN theory has been proposed to date, by \cite{2009_Roukema_et_al, 2020_Barrow}. As we will see in the following sections, this theory is unphysical and cannot be used to compute the backreaction.
The goal of this paper is to propose two NEN theories based on a minimal modification of the Newton-Cartan equation (proposed in \cite{1976_Kunzle}) and calculate the backreaction in both cases. This approach makes use of the results of \cite{2021_Vigneron} on the ability of the Newton-Cartan formalism to describe expansion as a fundamental field in Newton's theory. Especially, this approach is based on the 1+3-Newton-Cartan equations developed in that paper. We will not seek to justify one of these NEN theories with a non-relativistic limit of general relativity, which is the ultimate goal but is let for a futur work.

The preliminary section~\ref{sec::topo} defines the notion of ``(non)-Euclidean topology'' used in the present paper and some properties related to this notion. In section~\ref{sec::Strategy}, we introduce the strategy we use to define a NEN theory. Section~\ref{sec::Poisson_app} reviews the NEN theory of \cite{2009_Roukema_et_al, 2020_Barrow}. We give a precise mathematical definition of a NEN theory in section~\ref{sec::NEN_NC_app} based on the concept of Galilean manifolds. From this definition we propose two theories and calculate the backreaction in each case, along with discussing others of their properties. These results are summarised in table~\ref{tab:NEN}.

\section{What we mean by (non)-Euclidean topology}
\label{sec::topo}

In the literature on cosmology and general relativity, the terms ``Euclidean'' and ``non-Euclidean'' associated to ``geometry'' generally refer to the presence or not of a curvature. Our notion of Euclidean and non-Euclidean is different (more general) as it is associated to the topology of a manifold. 

The topology of \textit{closed} 3-manifolds is classified by Thurston's decomposition into eight distinct classes, each corresponding to a set of different (multi-connected) topologies having the same covering space. A 3-manifold $\Sigma$ has a \textit{Euclidean topology} if the covering space $\tilde{\Sigma}$ of $\Sigma$ is $\mE^3$. The topology in called \textit{non-Euclidean} if $\tilde{\Sigma} \not= \mE^3$. The most common non-Euclidean topologies, considered in the $\Lambda$CDM model, are the spherical and hyperbolic topologies having respectively $\tilde\Sigma = \mS^3$ and $\tilde\Sigma = \mH^3$. For the description of the other five types of topologies, see \citep[e.g.][]{1995_La_Lu}.

An important property relating the type of topology and the Ricci curvature tensor $\T \CR$ associated to any Riemannian metric $\T h$ defined on $\Sigma$ is:
\begin{align}
	\T\CR = 0 \ \Rightarrow \ \textrm{Euclidean topology}, \label{eq::property}
\end{align}
or equivalently,
\begin{align}
	\textrm{non-Euclidean topology} \ \Rightarrow \ \T\CR \not= 0. \nonumber
\end{align}
This is the reason why Newton's theory is defined on a Euclidean topology. This is also why a NEN theory requires a non-zero spatial Ricci tensor as discussed in the next section.

However, because the term ``non-Euclidean'' in the name ``non-Euclidean Newtonian theory'' refers first to the topology, the purpose of this modification of Newton's theory is not to take into account spatial curvature effects coming from general relativity, but effects of the global topology.

\section{The strategy: introduction of a spatial Ricci curvature}
\label{sec::Strategy}

The NEN theories we present in this paper are all constructed using the same strategy: we algebraically keep Newton's equations and we add a non-zero spatial Ricci curvature in the spatial connection such that the spatial topology is non-Euclidean. This is a heuristic approach as it is not yet justified from general relativity. Consequently there is a freedom on the choice of curvature, that we restrict with the following argument: the spatial topology in Newton’s theory is Euclidean because the spatial Ricci curvature $\TRt$ is zero. However, we see from property~\eqref{eq::property} that such a curvature tensor is not the only possible one in Euclidean topologies, i.e. in a 3-manifold with a Euclidean topology it is possible to define a metric whose Ricci curvature tensor is non-zero. Nevertheless, $\T\CR = 0$ is the simplest Ricci tensor we can define on a Euclidean topology. Therefore, it is natural to expect this tensor, in the case of a NEN theory, to be also the simplest one for the considered topology.

In this paper, we will be interested in either spherical or hyperbolic topologies. We will not consider the other five classes of topologies of the Thurston decomposition along with combinations of these topologies. Therefore, the simplest Ricci tensor is
\begin{equation}
	\TRt = \frac{\Rt(t)}{3} \T h, \label{eq::NEN_Ri}
\end{equation}
where $\T h$ is the spatial metric. We can see with the above formula that there is only one degree of freedom in $\T\CR$ given by the scalar curvature $\CR$, and because of the contracted Bianchi identity, $\CR$ is a spatial constant. Therefore, the only parameter on which $\CR$ can depend is the time $t$ present in the Newtonian theory. In the NEN theory of \citep{2009_Roukema_et_al,2020_Barrow} (section~\ref{sec::Poisson_app}), $t$ will parametrise a single 3-manifold (the spatial manifold) as in classical Newton's theory; in the NEN theories presented in section~\ref{sec::NEN_NC_app}, $t$ will have a geometrical origin as it will be linked to a foliation of a 4-manifold.

The main reasons to consider closed topologies either spherical or hyperbolic are:
\begin{enumerate}
	\item The closeness is the most physical choice of boundary conditions in a cosmological context. As we will see in section~\ref{sec::NEN2_constraint}, the closedness is also required in the calculation of the backreaction in our proposed NEN theories, along with the derivation of the Poisson equation.
	\item These topologies are globally isotropic, hence  respecting the cosmological principle.
\end{enumerate}

Another problem that we face with this heuristic strategy is related to the choice of equations on which we must add the curvature. Indeed, in Newton's theory the Poisson and the Raychaudhuri equations are equivalent. However, as pointed out by \citep{2022_Vigneron}, once we introduce a non-zero curvature in the spatial connection present in these equations, they are not equivalent anymore and therefore define two different NEN theories. In \cite{2009_Roukema_et_al, 2020_Barrow}, the authors use the Poisson equation. We detail this approach in the next section and we will see that physical inconsistencies appear. In section~\ref{sec::NEN_NC_app}, we will propose NEN theories based on the Newton-Cartan equation.\\


\section{Poisson equation approach}
\label{sec::Poisson_app}

The NEN theory of \cite{2009_Roukema_et_al,2020_Barrow} is defined with the following equations:
\begin{align}
	&\D_c g^c = - 4\pi G\rho \quad ; \quad \D_{[a}g_{b]} = 0 \quad ; \quad \Rt_{ab} = \frac{\Rt}{3} h_{ab},\label{eq::NEN_prop_ori} \\
	&\dot{\T v} = \T g + \T{a}_{\not=\mathrm{grav}} \quad ; \quad \dot\rho + \rho\D_cv^c = 0,
\end{align}
where $\T \D$ is the spatial connection related to the metric with the above Ricci curvature, $\T g$ is the gravitational field, $\rho$ the mass density, $\T v$ the velocity of the cosmic fluid, the dot derivative is the Lagrangian derivative with respect to $\T v$, and $\T{a}_{\not=\mathrm{grav}}$ is the non-gravitational spatial acceleration.

Throughout this paper, we denote indices running from 0 to 3 by Greek letters and indices running from 1 to 3 by Roman letters.

\subsection{The mass point gravitational field}
\label{sec::NEN_1_grav}

The authors in \cite{2009_Roukema_et_al,2020_Barrow} use this theory to compute the gravitational field created by one mass point. For the purpose of this section we only give the solution in the case where the topology\footnote{In the cases where the topology is multi-connected spherical, a sum over the holonomy group of the solution in $\mS^3$ gives the mass point solution (see \cite{2009_Roukema_et_al} for details).} is $\mS^3$ which can be found when solving equation~\eqref{eq::NEN_prop_ori} in spherical symmetry with $\rho = 0$. Then the gravitational field $\T g$ takes the form $g^a = \left(g^\xi, 0, 0\right)$ when using hyperspherical coordinates, in which the line element is
\begin{equation}
	\dd l^2 =  \Rc^2\left[\dd \xi^2 + \sin^2\left(\xi\right) \left(\dd \theta^2 + \sin^2\theta \dd \varphi \right)\right], \label{eq::NEN_hyper_coord}
\end{equation}
where $\xi \in [0, \pi]$, and $\Rc \coloneqq \sqrt{6/\Rt}$ is the curvature radius of the spherical space. In these coordinates, $r \coloneqq \Rc \, \xi$ corresponds to the distance to the origin $(0, 0, 0)$. Then we have
\begin{equation}
	g^\xi = \frac{A}{\Rc^2\sin^2\left(\xi\right)}, 
\end{equation}
with $A$ a constant. To be compatible with Newton's theory close to the center, i.e. close to the mass point, one must retrieve $g^\xi(\xi \sim 0) \sim -GM/r^2$. Thus we need to take $A = -G M$. We see that in the case of a spherical topology, the gravitational field diverges at the mass point position ($\xi = 0$) as expected, but also at the pole opposed to this position ($\xi = \pi$). The field around this pole is given by:
\begin{equation}
	g^\xi(\xi \sim \pi) = -\frac{GM}{\Rc^2\left(\pi - \xi\right)^2} + \mathcal{O}(1).
\end{equation}
This field points towards the origin, i.e. outwards from the pole $\xi = \pi$, which means that the latter pole behaves as a white hole.

\subsection{Problems of the approach}

The NEN theory defined above has two major problems which make it  unsuited for studying the backreaction problem or structure formation:
\begin{enumerate}
	\item The Poisson equation~\eqref{eq::NEN_prop_ori} requires that the total mass must be zero in a closed space. This can be derived when averaging the equation over the whole space, and using Stoke's theorem. In other words, the equation we solved is not the one with ``$\rho = M\delta(\xi)$'', but with $\rho = M[\delta(\xi) - \delta(\xi - \pi)]$. This explains why the mass point solution in spherical topology has two diverging points: one corresponding to an attractive (positive) mass and one to a repulsive (negative) mass, such that the sum of these two masses is zero. This solution is not physical as a spherical universe described by this NEN theory would be full of white holes.
	\item The proposed NEN theory cannot account for the spatial expansion, which, as shown in \cite{2021_Vigneron}, should be a fundamental field present in the equations. Furthermore introducing an expansion velocity of the form ${v}^a_\mathrm{H} = H(t)x^a$ as in \cite{1980_Peebles}, is not possible because this construction requires the use of Cartesian coordinates, only defined for Euclidean topologies.
\end{enumerate}

In the approach of \cite{2009_Roukema_et_al,2020_Barrow}, the Poisson equation is algebraically kept and thus is considered as a fundamental feature of a Newtonian theory regardless of the topology. Another approach would be to consider that the fundamental feature which should be common to Newton's and NEN theories is the Galilean invariance. This is the approach we followed and which is detailed in the next section.

\section{The Newton-Cartan approach}
\label{sec::NEN_NC_app}

\subsection{Definition of a NEN theory}
\label{sec::def_NEN}

In \cite{2022_Vigneron} and in the introduction of the present paper we defined a NEN theory as a theory whose equations should reduce to Newton's equations on scales small with respect to the size of the closed space, which corresponds to the spatial curvature scale in our strategy. However it is not clear to us how to define the mathematical procedure that would test if the theory is ``locally Newtonian'' or not. For the NEN theory proposed in section~\ref{sec::Poisson_app}, we can only say that if we assume $\T \Rt \sim \T 0$, we retrieve Newton's equations. In what follows, we propose a better definition which is based on a mathematical framework that keeps the essence of Newton's theory, the Galilean invariance. 

In \cite{2021_Vigneron}, we showed that Newton's theory is best formulated, and argued that it must be, in its geometrised form, i.e. the Newton-Cartan theory. In particular, we showed that from this formulation, the expansion arises as an emerging field of the theory. In this formulation, the equations are defined on a 4-manifold whose structure is invariant under local Galilean transformations (contrary to general relativity where the structure is invariant under local Lorentzian transformations). Therefore, in this formulation the Galilean invariance of spacetime is taken as a fundamental principle. We can therefore redefine a NEN theory as follows:\saut
\begin{mdframed}
\definition{A \textbf{non-Euclidean Newtonian theory} is a theory defined on a Galilean manifold whose spatial sections have a non-Euclidean topology. It is given by equations relating the Riemann tensor of the Galilean structure to the energy content of the manifold.}
\end{mdframed}

From this definition and the strategy proposed in section~\ref{sec::Strategy}, it seems natural that to define a NEN theory, one should add the non-zero spatial curvature directly in the Newton-Cartan equations. This was actually proposed by \cite{1976_Kunzle}, however the expansion law, along with the full 3-dimensional system of equations, were not derived in that paper. This is done in the next sections, but we first recall the formalism behind the Newton-Cartan theory.

\subsection{The Newton-Cartan theory}


A \textit{Galilean manifold} is a differentiable 4-manifold $\CM$ equipped with a \textit{Galilean structure} $(\T\tau, \T h, \T\nabla)$, where $\T\tau$ is an exact 1-form, $\T h$ is a symmetric (2,0)-tensor of rank 3, with $h^{\alpha\mu}\tau_\mu = 0$, and $\T\nabla$ is a connection compatible with $\T\tau$ and $\T h$, called a \textit{Galilean connection}:
\begin{equation}
	\nabla_\alpha \tau_\beta = 0 \quad ; \quad \nabla_\gamma h^{\alpha\beta} = 0. \label{eq::NC_def_structure}
\end{equation}
A vector $\T u$ is called $\textit{a unit timelike vector}$ if $u^\mu\tau_\mu = 1$, and an (n,0)-tensor $\T T$ is called \textit{spatial} if $\tau_\mu {T^{... \overset{\overset{\alpha}{\downarrow}}{\mu} ...}} = 0$ for all $\alpha \in \llbracket1,n\rrbracket$. The exact 1-form $\T \tau$ defines a foliation $\folGR$ in $\CM$, where $\Sigma_t$ are spatial hypersurfaces in $\CM$ defined as the level surfaces of the scalar field $t$, with $\T \tau = \T\dd t$. The time metric $\T \tau$ and the space metric $\T h$ define what we can call a preferred foliation in $\CM$, i.e. preferred (or absolute) space and time. This is the Newtonian picture of spacetime. However, from the definition alone of the Galilean structure, no constraint exists on the spatial sections $\Sigma_t$, especially on the topology of these sections and their curvature related to the spatial metric $\T h$. Such constraints will only come from the physical equations relating the geometry of the Galilean manifold with its energy content.

From the coefficients $\Gamma^\gamma_{\alpha\beta}$ of the Galilean connection, one can define a Riemann tensor as ${\Rf^{\sigma}}_{\alpha\beta\gamma} \coloneqq 2 \, \partial_{[\beta}\Gamma^\sigma_{\gamma]\alpha} + 2 \,  \Gamma^\sigma_{\mu[\beta} \Gamma^\mu_{\gamma]\alpha}$. Then the Newton-Cartan equations are:
\begin{align}
	\nabla_\mu T^{\mu\alpha} &=0, \label{eq::NC_Conservation} \\
	\Rf_{\alpha\beta} &= \tau_\alpha \tau_\beta \left( 4\pi G \tau_\mu \tau_\nu T^{\mu\nu} - \Lambda \right), \label{eq::NC_Einstein} \\
	h^{\mu[\alpha}{\Rf^{\beta]}}_{(\gamma\sigma)\mu} &= 0, \label{eq::NC_Kunzle}
\end{align}
where $\Lambda$ is the cosmological constant and $\T T$ is symmetric and corresponds to the energy-momentum tensor of the matter.

Projecting twice the second equation along the spatial metric leads to $\TRt = 0$, where $\TRt$ is the Ricci tensor related to the spatial metric. This means that the Newton-Cartan equations impose the topology of the sections $\Sigma_t$ to be Euclidean. Therefore allowing for non-Euclidean topologies requires a modification of these equations.\\

\subsection{Some remarks about Galilean structures}

An important technical point about Galilean structures is that that they do not introduce any object, like a Lorentzian metric, that would induce an isomorphism between forms and vectors on $\CM$. Therefore, raising or lowering indices is not possible. That is why all the spacetime tensors we introduce in this paper will always be written with the same types of indices i.e. contravariant/covariant. For more details regarding mathematical properties of Galilean structures see \citep{1972_Kunzle, 2021_Vigneron}.

It is still possible to relate Galilean structures with Lorentzian structures used in general relativity. This has first been done in \citep{1976_Kunzle}, where the author defines the Newtonian limit using these structures: in this limit, the Lorentzian structure of general relativity tends to a Galilean structure. In other words, the Newtonian limit transforms a Lorentz invariant theory into a Galilean invariant theory, and for this reason is also called \textit{Galilean limit} (see \citep{2022_Vigneron_c} for a detailed study of the relation between this limit of global topology).

Finally, the Galilean invariance at the origin of the definition of Galilean structures is defined as a \textit{local} invariance between frames (see \cite{1972_Kunzle}), whereas a global invariance would necessarily require a flat Euclidean space. That is why Galilean manifolds can have non-zero spatial curvature and non-Euclidean topologies for their spatial sections.

\subsection{A curvature in Newton-Cartan}

To allow for $\TRt \not= 0$, \cite{1976_Kunzle} proposed\footnote{K\"unzle claims to give references for the modification~\eqref{eq::NEN_Einstein_R}, but these are unrelated to this equation. We can therefore consider \cite{1976_Kunzle} to be the first occurrence of this modification.} to modify the system~\eqref{eq::NC_Conservation}-\eqref{eq::NC_Kunzle} by adding a `curvature term' in the second equation, leading to the system:
\begin{align}
	\nabla_\mu T^{\mu\alpha} &=0, \label{eq::NEN_Conservation} \\
	\Rf_{\alpha\beta} - \frac{\Rt}{3}\bb{B}_{\alpha\beta} &= \tau_\alpha \tau_\beta \left( 4\pi G \tau_\mu \tau_\nu T^{\mu\nu} - \Lambda \right), \label{eq::NEN_Einstein_R}\\
	h^{\mu[\alpha}{\Rf^{\beta]}}_{(\gamma\sigma)\mu} &= 0, \label{eq::NEN_Kunzle}
\end{align}
where $\T B$ is a unit timelike vector and $\bb{B}_{\alpha\beta}$ is the projector orthogonal to this vector, defined as:
\begin{equation}
	\bbB_{\alpha\mu} B^\mu \coloneqq 0 \quad ; \quad \bbB_{\alpha\mu}h^{\mu\beta} \coloneqq {\delta_\alpha}^\beta - \tau_\alpha B^\beta. \label{eq::NC_def_bb}
\end{equation}
From now we also assume that $\T T$ corresponds to the energy-momentum tensor of a general fluid with
\begin{equation}
	T^{\alpha\beta} \coloneqq \rho u^\alpha u^\beta + p h^{\alpha\beta} + 2q^{(\alpha}u^{\beta)} + \pi^{\alpha\beta}, \label{eq::T^u}
\end{equation}
where $\T u$ is timelike and is the 4-velocity of the fluid, $p$ its pressure, $\T q$ its heat flux and $\T\pi$ its anisotropic stress.

When projected twice along $h^{ab}$, equation~\eqref{eq::NEN_Einstein_R} indeed leads to $\Rt^{ab} = \frac{\Rt}{3}h^{ab}$. This modification of the Newton-Cartan equations is the simplest one allowing for a non-zero spatial Ricci curvature, and in the sense given in section~\ref{sec::def_NEN}, the system~\eqref{eq::NEN_Conservation}-\eqref{eq::NEN_Kunzle} defines a NEN theory.

In the framework of Galilean manifolds, any orthogonal projector $\bb{B}_{\alpha\beta}$ can play the role of a twice covariant spatial metric. This is because in any coordinate system adapted to the foliation  $\folGR$ we have $\bb{B}_{ab} = h_{ab}$, with $h_{ab}$ the inverse matrix of $h^{ab}$. That is why the additional term in equation~\eqref{eq::NEN_Einstein_R} features $\bb{B}_{\alpha\beta}$. As a consequence there is a freedom on the choice of the vector $\T B$, i.e. on the choice of observer to which the curvature term is related. 

We can choose $\T B$ to be a `preferred' unit timelike vector. Following \cite{2021_Vigneron}, two such vectors can be chosen:
\begin{enumerate}
	\item $\T B = \T u$: the curvature term is related to the fluid observer.
	\item $\T B = \T G$: the curvature term is related to the Galilean observer. It is defined such that its tilt with respect to the fluid $\T v = \T u - \T G$ is the one present in the Scalar-Vector-Tensor (SVT) decomposition of the expansion tensor $\T\Theta$ of $\T u$, defined in equation~\eqref{eq::NC_decomp} (see \cite{2021_Vigneron} for a detailed description of the Galilean vector). In section~\ref{sec::bite} we will provide a better definition for $\T G$ based on the results of section~\ref{sec::chatte}.
\end{enumerate}
In the following sections we will develop the NEN theory in both cases: we will derive the 1+3-system, or kinematical system (similarly as in \cite{2021_Vigneron}), and the gravitational system, i.e. featuring the gravitational field and the expansion law.  The cosmological backreaction will be given, thus generalising the {\BET} in both cases. Finally we will show whether or not each theory is suited for $N$-body calculations.

\subsection{NEN theory I: Fluid observer curvature term}
\label{sec::NEN_I}

In this section, we consider $\T B \coloneqq \T u$. 

\subsubsection{The system of equations.}
\label{sec::caca}

The derivation of the 1+3-system of equations is done by projecting the modified Newton-Cartan equations with respect to $\T u$ and $\T h$. We will not detail its full derivation, which is similar to the one made by \cite{2021_Vigneron}. The main difference comes from the projections of the additional curvature term in the Newton-Cartan equation. We have
\begin{equation}
\frac{\Rt}{3}\bb{u}_{\mu\nu}h^{a\mu}h^{b\nu} = \frac{\Rt}{3} h^{ab} \quad ; \quad \frac{\Rt}{3}\bb{u}_{\mu\nu}u^{\mu}h^{a\nu} = 0 \quad ; \quad \frac{\Rt}{3}\bb{u}_{\mu\nu}u^\mu u^\nu = 0.
\end{equation}
Then when introducing the expansion tensor $\Theta^{\alpha\beta} \coloneqq h^{\mu(\alpha} \nabla_\mu u^{\beta)}$ with its trace  ${\theta \coloneqq \Theta^{\mu\nu} \bb{u}_{\mu\nu}} = \nabla_\mu u^\mu$ and the vorticity tensor $\Omega^{\alpha\beta} \coloneqq h^{\mu[\alpha} \nabla_\mu u^{\beta]}$ of the fluid, one gets the following 1+3-system, or kinematical system, composed of: evolution equations
\begin{align}
	&\left(\partial_t - \Lie{\T \beta}\right)\rho	= - \rho \theta - \D_c q^c, \label{eq::NEN2_conser_1b} \\
	&\left(\partial_t - \Lie{\T \beta}\right)h^{ab}	= -2 \Theta^{ab}, \label{eq::NEN2_Theta_hb} \\
	&\left(\partial_t - \Lie{\T \beta}\right) \theta	= -4\pi G \rho + \Lambda + \D_c \Accu^c - \Theta^{cd} \Theta_{cd} + \Omega^{cd} \Omega^{cd}, \label{eq::NEN2_Einstein_ab}  \\
	&\left(\partial_t - \Lie{\T \beta}\right) \Omega_{ab}	= \D_{[a} \Accu_{b]}, \label{eq::NEN2_vorticity_b}
\end{align}
and constraint equations
\begin{align}
	&\D_c \left(\Theta^{ac} + \Omega^{ac}\right) - \D^a \theta = 0, \label{eq::NEN2_Mom} \\
	&\D_{[a} \Omega_{bc]}	= 0, \label{eq::NEN2_Gauss_Omega} \\
	&{\Rt^{ab}= \color{Mred} \frac{\Rt}{3} h^{ab}}, \color{black} \label{eq::NEN2_Einstein_3b} \\
	&\Rt^{d[abc]}			= 0, \label{eq::NEN2_Kunzle_3b}
\end{align}
with 
\begin{align}
	\rho \, \Accu^a	&= -\D^a P -\D_c \pi^{ca} - \left[\left(\partial_t - \Lie{\T \beta}\right) q^a + q^a \theta + 2 q_c \left(\Theta^{ca} + \Omega^{ca}\right)\right]. \label{eq::NEN2_conser_2b}
\end{align}
$\T \beta$ is the shift vector, it is spatial and corresponds to the choice of adapted coordinate system with $\T \beta \coloneqq \T\partial_t - \T u$, and $(\T\partial_t)^\mu \tau_\mu \coloneqq 1$ (see section II.C.3 in \cite{2021_Vigneron} for more details on coordinate choices in the Newton-Cartan formalism). The 4-acceleration $\tensor[^{\T u}]{\T a}{}$ of the fluid plays the role of a non-gravitational 3-acceleration, and from now on will be denoted as $\T{a}_{\not= \rm grav}$. Note that centrifugal and Coriolis accelerations are not part of $\T{a}_{\not= \rm grav}$. These accelerations arise from a non-Galilean choice of coordinates and are contained in the Lie derivative term $\Lie{\T\beta}v^a$ in equation~\eqref{eq::NC_def_g} (also see section~II. in \cite{2020_Vigneron} for details on how these accelerations can be described by the shift vector).

The only difference of the system~\eqref{eq::NEN2_conser_1b}-\eqref{eq::NEN2_conser_2b} with respect to the 1+3-Newton-Cartan system (equations~(36)-(44) in \cite{2021_Vigneron}) is the curvature equation~\eqref{eq::NEN2_Einstein_3b}.\\

\remark{If we want to retrieve the mass conservation equation we need to assume $\T q = 0$ in equation~\eqref{eq::NEN2_conser_1b}. This was also the case in Newton-Cartan as shown by \cite{1976_Kunzle}.}


\subsubsection{Solving the constraint equations.}
\label{sec::NEN2_constraint}

A key step in the derivation of the gravitational system, is the resolution of the constraint equations~\eqref{eq::NEN2_Mom} and \eqref{eq::NEN2_Gauss_Omega}. But while these equations are algebraically the same as in the Euclidean case, because the spatial curvature is not zero anymore, their solution is different.

The constraint~\eqref{eq::NEN2_Gauss_Omega} implies that $\T \Omega$ can be written as
\begin{equation}
	\Omega_{ab} = \D_{[a}w_{b]} + \rot_{ab},
\end{equation}
with $\T\rot$ a harmonic 2-form and $\T w$ a spatial vector. In the case of a spherical topology, $\T\omega$ is necessarily zero. Because we consider a closed 3-manifold, we can decompose the expansion tensor $\T\Theta$ into scalar, vector and tensor parts \citep{1973_York}:
\begin{equation}
	\Theta_{ab} = \chi h_{ab} + \D_{(a}v_{b)} + \Xi_{ab}, \label{eq::NC_decomp}
\end{equation}
with $\chi$ a scalar field, $\T v$ a spatial vector and $\T \Xi$ is a transverse-traceless tensor , i.e. ${\Xi_c}^c \coloneqq 0$ and $\D_c\Xi^{ca} \coloneqq 0$. Then, from equation~\eqref{eq::NC_decomp} we have $\theta = 3\chi + D_c v^c$, and the momentum constraint~\eqref{eq::NEN2_Mom} becomes
\begin{equation}
	- 2\D_a\chi + \D^c\DOm{_a_c} + v_a \frac{\Rt}{3} = 0,
\end{equation}
where $\DOm{_a_b} \coloneqq \D_{[a} w_{b]} -  \D_{[a} v_{b]}$. We introduce the Hodge decomposition on $\T v$ as ${\T v \eqqcolon \T\D \Phi_v + \textrm{curl}\T A + \T\lambda}$, where $\Phi_v$ and $\T A$ are respectively the scalar potential and the vector potential of $\T v$, and $\T\lambda$ is a harmonic 1-form. As the term $\D_a\chi$ is vorticity-free and the term $\D^c\DOm{_c_a}$ is divergence-free, we have
\begin{align}
	&\D_a\chi = \D_a\left(\frac{\Rt}{6}\Phi_v\right), \label{eq::NEN2_chi_phi} \\
	&\D^c\DOm{_c_a} = \frac{\Rt}{3} \left(\epsilon_{acd}\D^c A^d + \lambda_a\right). \label{eq::NEN2_v_w}
\end{align}
We see that, contrary to the Euclidean case in \cite{2021_Vigneron}, the first equation~\eqref{eq::NEN2_chi_phi} does not imply anymore that $\chi$ is only a function of time. Instead we have
\begin{equation}
	\chi(t,x^i) = cst(t) + \frac{\Rt(t)}{6}\Phi_v(t,x^i).
\end{equation}
The second equation~\eqref{eq::NEN2_v_w} shows that $\T v \not= \T w$ in general. This is a radical difference with the Euclidean case: the vectors inside the expansion and vorticity tensors are different. We finally have
\begin{equation}
	\Theta_{ab} = \chi(t,x^i) h_{ab} + \D_{(a}v_{b)} + \Xi_{ab} \quad ; \quad \Omega_{ab} = D_{[c}v_{a]} + \DOm{_a_b} + \rot_{ab}. \label{eq::NEN2_Theta_Omega_sol}
\end{equation}

\subsubsection{The gravitational system.}

The system~\eqref{eq::NEN2_conser_1b}--\eqref{eq::NEN2_conser_2b} of the present NEN theory can be rewritten to feature the gravitational field and the analogue to the Poisson equation. The gravitational field is defined as the opposite to the 4-acceleration of the Galilean observer (see section~V in \cite{2021_Vigneron}) and is given by
\begin{equation}
	g^a \coloneqq \left(\partial_t - \Lie{\T \beta}\right) v^a + 2v^c\left({\Theta_c}^a + {\Omega_c}^a\right) - v^c\D_c v^a - \Accu^a. \label{eq::NC_def_g}
\end{equation}
Then, accounting for the expression of $\T\Theta$ and $\T\Omega$ in equation~\eqref{eq::NEN2_Theta_Omega_sol} and defining the tensor $\T\GM \coloneqq \TDOm + \T\rot$, the gravitational system takes the form:\saut

\begin{mdframed}
\underline{Non-Euclidean Newtonian theory I:} The gravitational field is given by
\begin{align}
	g^a			&= \left(\partial_t - \Lie{\T \beta}\right) v^a + v^c\D_c v^a+ 2v^c \left(\chi {\delta_c}^a + {\Xi_c}^a + {\GM_c}^a\right) - (a_{\not= \rm grav})^a, \label{eq::NEN2_def_g_1} \\
	\D_c g^c		&= -4\pi G {\rho} - {\Xi_{cd}\Xi^{cd}} + {\GM_{cd}\GM^{cd}} - \frac{\Rt}{3}{v_cv^c} \label{eq::NEN2_New_Gaus_1} \\
				&\qquad - 3\left[{\left(\partial_t - \Lie{\T \beta}\right)\chi} + {\chi^2} - {v^c\D_c\chi}\right],  \nonumber \\
	\D_{[a} g_{b]}	&= -\left(\partial_t - \Lie{\T \beta + \T v}\right)\GM_{ab}, \label{eq::NEN2_Faraday}
\end{align}
with
\begin{align}
	\chi &= cst(t) + \frac{\Rt(t)}{6}\Phi_v, \\
	D^c\GM_{ca} &= \frac{\Rt}{3} \left(V_a - \D_a\Phi_v\right). \label{eq::NEN2_Amper}
\end{align}
where $\Phi_v$ is the scalar potential of $\T v$, and $\T \GM$ a closed 2-form. These equations are completed by
\begin{align}
	\Rt_{ab}	&= \frac{\Rt(t)}{3} h_{ab}, \label{eq::NEN2_sys_Ri} \\
	\left(\partial_t - \Lie{\T \beta}\right)h_{ab}	&= 2 \left(\chi h_{ab} + \D_{(a}v_{b)} + \Xi_{ab}\right), \label{eq::NEN2_Theta_hb_cosmo} \\
	\left(\partial_t - \Lie{\T \beta}\right) \rho &= -\rho\left(3H + \D_c v^c\right),
\end{align}
and the averaged equations
\begin{align}
	3\left(\dot H + H^2\right) + 4\pi G \Saverage{\rho} - \Lambda = \CQ_\Sigma \quad ; \quad \label{NEN2_exp_law}\Saverage{\rho} = \frac{M_{\rm tot}}{V_{\Sigma}(t)},
\end{align}
where $V_\Sigma$ is the volume of $\Sigma$ and $H(t) \coloneqq \dot{V}_\Sigma/3V_\Sigma = \Saverage{\chi}$ its global expansion rate, and where
\begin{align}
	\CQ_\Sigma = - \Saverage{\Xi_{cd}\Xi^{cd}} + \Saverage{\GM_{cd}\GM^{cd}} - \frac{\Rt}{3}\Saverage{v_cv^c} + \frac{\Rt}{6}\left(\Saverage{\Phi_v^2} - \Saverage{\Phi_v}^2\right), \label{eq::NEN2_Back}
\end{align}
where the average on a scalar $\psi$ over the whole spatial manifold is defined as $\average{\psi}{\PerD}(t) \coloneqq \frac{1}{V_\PerD}\int_\PerD \psi \sqrt{\mathrm{det}(h_{ab})}\dd^3 x$.
\end{mdframed}\saut

\subsubsection{Gravitomagnetism.}

In the above system we introduced the closed 2-form $\T\GM \coloneqq \TDOm + \T\rot$. This field corresponds to a gravitomagnetic field whose Newton-Faraday equation is equation~\eqref{eq::NEN2_Faraday} and whose Newton-Amp\`ere equation is equation~\eqref{eq::NEN2_Amper}. The later does not feature any displacement current. The analogue to the current density is $J_a \coloneqq \frac{\Rt}{3} \left(V_a - \D_a\Phi_v\right)$.

We recall that the ``non-Euclidean'' in the name NEN theory refers to the topology and not the spatial curvature. Therefore the NEN theory is an extension of Newton's theory whose purpose is not to take into account post-Newtonian, i.e. relativistic, effects coming from spatial curvature, but effects coming from the global topology and which should be present at the zeroth order of a non-relativistic limit of general relativity. In this sense, we expect, as for classical Newton's theory, that no-gravitomagnetism should be present in the ``right'' NEN theory, i.e. $D_{[a}g_{b]} = 0$. This phenomenon is a relativistic effect that should appear only when considering the first order of the limit (e.g. \citet{1995_Kofman_et_al}). Gravitomagnetism will not be present in the second NEN theory presented in section~\ref{sec::NEN_II}.

The reason why topology is not something relativistic is because this is a property of the 4-manifold $\CM$ we consider for our Universe, which does not depend on the structure we define on this manifold: i.e. whether $\CM$ is equipped with a Lorentzian or a Galilean structure. This implies that topology should not only be considered under the framework of a relativistic theory, but also under the framework of a Newtonian (i.e. Galilean invariant) theory, as done in the present paper.

\subsubsection{Expansion law and generalised \BET.}
\label{sec::NEN_back}

The term $\CQ_\Sigma$ in the expansion law~\eqref{NEN2_exp_law} measures the difference of the global expansion of the manifold $\Sigma$ with respect to the Friedmanian case. It is called the cosmological backreaction. With respect to the Euclidean case (derived in \cite{1997_Buchert_et_al,2021_Vigneron}) where `$\CQ_\Sigma = - \Saverage{\Xi_{cd}\Xi^{cd}} + \Saverage{\rot_{cd}\rot^{cd}}$', equation~\eqref{eq::NEN2_Back} features additional terms. These terms depend on the scalar curvature, the velocity $\T v$, and the gravitomagnetic field $\T\GM$. This shows that the backreaction depends on the class of topology, via the sign of $\Rt$, but also on the dynamical properties of the fluid, via $\T v$ and $\T \GM$. Especially the term `$-\frac{\CR(t)}{3}\left\langle v^cv_c \right\rangle_{\Sigma}$' shows that the mean specific energy of the fluid directly affects the backreaction. We expect this specific energy to be more important in the late Universe when structure formation increases. Then, formula~\eqref{eq::NEN2_Back} suggests that the structure formation might play a major role on the global expansion of our Universe if the topology is non-Euclidean. This behaviour is also present in general relativity \cite{2020_Brunswic_et_al}, but in our case it arises at a non-relativistic level, as it shows that the Universe might have a local Newtonian dynamics, but a global dynamics which differs from the one of a homogeneous Universe. However, quantifying the mean specific energy remains difficult because it depends on the mesoscopic scale, which is not well defined in cosmology.

\subsubsection{Space--time separation of the spatial metric?}

In important feature of Newton's theory is the fact that the spatial metric can be separated in space and time (if the expansion is isotropic) with ``$h_{ab}(t,x^i) = a^2(t) \tilde{h}_{ab}(x^i)$'', where $\bar{h}_{ab}(x^i)$ is a flat metric. This separation ensures that the spatial metric has no local dynamics and that the resolution of Newton's equations is ``simple''. Such a separation is not possible in the proposed NEN theory because $\chi$ depends on time and space. Instead, when solving equation~\eqref{eq::NEN2_Theta_hb_cosmo} with $\T\Xi = 0$ we can only have the following conformal form (which is valid in the Galilean coordinates: $\T\beta = -\T v$):
\begin{equation}
	h_{ab}(t,x^i) = \psi^2(t, x^i) \bar{h}_{ab}(x^i), \label{eq::NEN2_sepa_met_2}
\end{equation}
where $\partial_t\psi/\psi \coloneqq \chi$. We can still define a background metric $\bar{\T h}$ independent of time (as in the Euclidean case), but its Ricci curvature tensor $\T{\bar{\Rt}}$ has a traceless part, and its scalar curvature ${\bar{\Rt}}$ is not spatially constant anymore (see the conformal relation~(7.42) in \cite{2012_GG}). Such a feature makes the equations hardly tractable, which is something we do not expect for a Newtonian-like theory. The only way to have a complete separation is to suppose $\T\D\Phi_v = 0$. But in that case the vector $\T v$, which is interpreted as the spatial velocity of the fluid, is purely rotational.

In conclusion, it is difficult to consider that the proposed NEN theory is physically relevant.

\subsubsection{Limitations of this NEN theory.}
\label{sec::NEN1_limitations}

We list here the features of this theory which, to our opinion, make it unphysical as a Newtonian theory:
\begin{enumerate}
	\item There is a gravitomagnetic field which cannot be taken to be zero without loss of generality on $\T v$.
	\item The vectors inside $\T\Theta$ and $\T\Omega$ are different, i.e. $\DOm{_a_b} \not= 0$.
	\item The spatial metric cannot be separated in space and time without loss of generality on $\T v$. No ``simple'' background metric can be defined, and the system of equations is hardly tractable.
	\item There is no exact $N$-body description because the ``Poisson equation''~\eqref{eq::NEN2_New_Gaus_1} is not linear and does not only feature the gravitational field, but also the velocity. In other words, the gravitational field created by two mass points (described by two Dirac fields in the density field) does not corresponds to the sum of the gravitational field solution of~\eqref{eq::NEN2_New_Gaus_1} for a single Dirac field, and this mainly because of the presence of $\frac{\Rt}{3}{v_cv^c} $.
\end{enumerate}
Furthermore, as the curvature term in the modified Newton-Cartan equations is added for topological reasons, we do not want to interpret it as a curvature-fluid coupling term. Therefore, in the absence of a fluid we would expect this term to be still present, which seems difficult to concile with the fact that $\frac{\Rt}{3}\bb{u}_{\alpha\beta}$ explicitly depends on the presence of a fluid.

\subsection{NEN theory II: Galilean observer curvature term}
\label{sec::NEN_II}

In this section, we consider $\T B \coloneqq \T G$. 

\subsubsection{The system of equations.}

The curvature term projected with respect to $\T u$ and $\T h$ gives
\begin{equation}
	\frac{\Rt}{3}\bb{G}_{\mu\nu}h^{a\mu}h^{b\nu} = \frac{\Rt}{3} h^{ab} \quad ; \quad \frac{\Rt}{3}\bb{G}_{\mu\nu}u^{\mu}h^{a\nu} = \frac{\Rt}{3}v^a \quad ; \quad \frac{\Rt}{3}\bb{G}_{\mu\nu}u^\mu u^\nu = \frac{\Rt}{3}v^\mu v^\nu \bb{G}_{\mu\nu}.
\end{equation}
Then the 1+3-system of equations is composed of: evolution equations
\begin{align}
	&\left(\partial_t - \Lie{\T \beta}\right)\rho	= - \rho \theta - \D_c q^c, \label{eq::NEN3_conser_1b} \\
	&\left(\partial_t - \Lie{\T \beta}\right)h^{ab}	= -2 \Theta^{ab}, \label{eq::NEN3_Theta_hb} \\
	&\left(\partial_t - \Lie{\T \beta}\right) \theta	= -4\pi G \rho + \Lambda + \D_c \Accu^c - \Theta^{cd} \Theta_{cd} + \Omega^{cd} \Omega^{cd} \, {\color{Mred} - \frac{\Rt}{3} v^cv_c}, \label{eq::NEN3_Einstein_ab} \\
	&\left(\partial_t - \Lie{\T \beta}\right) \Omega_{ab}	= \D_{[a} \Accu_{b]}, \label{eq::NEN3_vorticity_b}
\end{align}
and constraint equations
\begin{align}
	& {\D_c \left(\Theta^{ac} + \Omega^{ac}\right) - \D^a \theta = \color{Mred} \frac{\Rt}{3} v_a}, \label{eq::NEN3_Mom} \\
	&\D_{[a} \Omega_{bc]}	= 0, \label{eq::NEN3_Gauss_Omega} \\
	& {\Rt^{ab}= \color{Mred}\frac{\Rt}{3} h^{ab}}, \label{eq::NEN3_Einstein_3b} \\
	&\Rt^{d[abc]}			= 0, \label{eq::NEN3_Kunzle_3b}
\end{align}
with 
\begin{align}
	\rho \, \Accu^a			&= -\D^a P -\D_c \pi^{ca} - \left[\left(\partial_t - \Lie{\T \beta}\right) q^a + q^a \theta + 2 q_c \left(\Theta^{ca} + \Omega^{ca}\right)\right]. \label{eq::NEN3_conser_2b}
\end{align}

With respect to the first proposed NEN theory in section~\ref{sec::NEN_I}, this system of equations features additional terms in the Raychaudhuri equation~\eqref{eq::NEN3_Einstein_ab} and in the momentum constraint~\eqref{eq::NEN3_Mom}. These terms will radically change the solutions to the constraint equations and the expansion law, as presented in the following sections.

\subsubsection{Solving the constraint equations.}

Following the same logic than in section~\ref{sec::NEN2_constraint} for the first NEN theory, the momentum constraint~\eqref{eq::NEN3_Mom} becomes
\begin{equation}
	- 2\D_a\chi + \D^c\DOm{_a_c} + v_a \frac{\Rt}{3} = v_a \frac{\Rt}{3},
\end{equation}
which implies through uniqueness of the Hodge decomposition
\begin{align}
	&\D_a\chi = 0, \label{eq::NEN3_chi_phi} \\
	&\D^c\DOm{_c_a} = 0. \label{eq::NEN3_v_w}
\end{align}
This is equivalent to the Euclidean case in \cite{2021_Vigneron}. Therefore $\chi$ depends only on time, and $\TDOm = 0$. We finally have
\begin{equation}
	\Theta_{ab} = H(t) h_{ab} + \D_{(a}v_{b)} + \Xi_{ab} \quad ; \quad \Omega_{ab} = D_{[c}v_{a]} + \rot_{ab},\label{eq::NEN3_Theta_Omega_sol}
\end{equation}
where $H \coloneqq \dot{V}_\Sigma/3V_\Sigma = \Saverage{\chi} = \chi$ is the expansion rate of the total volume $V_\Sigma$ of the spatial manifold $\Sigma$.

\subsubsection{The gravitational system.}

The gravitational system is obtained when rewriting the system~\eqref{eq::NEN3_conser_1b}--\eqref{eq::NEN3_conser_2b} as function of the gravitational field, and accounting for the expressions of $\T\Theta$ and $\T\Omega$ in equation~\eqref{eq::NEN3_Theta_Omega_sol}:\saut

\begin{mdframed}
\underline{Non-Euclidean Newtonian theory II:} The gravitational field is given by
\begin{align}
	g^a &= \left(\partial_t - \Lie{\T \beta}\right) v^a + v^c\D_c v^a+ 2v^c \left(H {\delta_c}^a + {\Xi_c}^a + {\omega_c}^a\right) - (a_{\not= \rm grav})^a, \label{eq::NEN_def_g_1} \\
	\D_c g^c &= -4\pi G \widehat{\rho} - \widehat{\Xi_{cd}\Xi^{cd}} + \widehat{\rot_{cd}\rot^{cd}}, \label{eq::NEN_New_Gaus_1} \\
	\D_{[a} g_{b]} &= -\left(\partial_t - \Lie{\T \beta + \T v}\right)\rot_{ab},
\end{align}
where $\widehat{\psi} \coloneqq \psi - \Saverage{\psi}$. These equations are completed by
\begin{align}
	\Rt_{ab} &= \frac{\Rt(t)}{3} h_{ab}, \label{eq::NEN_sys_Ri} \\
	\left(\partial_t - \Lie{\T \beta}\right)h_{ab}	&= 2 \left(H h_{ab} + \D_{(a}v_{b)} + \Xi_{ab}\right), \label{eq::NEN_Theta_hb} \\
	\left(\partial_t - \Lie{\T \beta}\right) \rho &= -\rho\left(3H + \D_c v^c\right),
\end{align}
and the averaged equations
\begin{align}
	3\left(\dot H + H^2\right) + 4\pi G \Saverage{\rho} - \Lambda = - \Saverage{\Xi_{cd}\Xi^{cd}} + \Saverage{\rot_{cd}\rot^{cd}} \quad; \quad  \Saverage{\rho} = \frac{M_{\rm tot}}{V_{\Sigma}(t)}. \label{eq::NEN_sys_av_rho}
\end{align}
\end{mdframed}

A remarkable thing is that the above system of equations is algebraically equivalent to Newton's gravitational system of equations, but with a non-zero curvature in the spatial connection. Especially, contrary to the 1+3-system~\eqref{eq::NEN3_conser_1b}-\eqref{eq::NEN3_conser_2b} and the spacetime system~\eqref{eq::NEN_Conservation}-\eqref{eq::NEN_Kunzle}, the gravitational system~\eqref{eq::NEN_def_g_1}-\eqref{eq::NEN_sys_av_rho} does not feature any additional curvature terms with respect to the Euclidean case.\\

\remark{This system of equations is similar to the one proposed by \cite{2009_Roukema_et_al,2020_Barrow}, the main difference being that expansion is present and that the Poisson equation~\eqref{eq::NEN_New_Gaus_1} appears in its cosmological form, i.e. featuring $\widehat{\rho}$ instead of $\rho$. Actually, this form should be considered as the right form of the Poisson equation, the one with just $\rho$ being a restricted case when we consider an isolated system in an infinite space, implying $\Saverage{\rho} = 0$ and $\widehat{\rho} = \rho$.}

\subsubsection{Expansion law or generalised \BET.}

The expansion law is the same as for Newton's theory. Therefore, the same conclusions apply: the expansion law for ${\T\Xi = 0 = \T\rot}$ is given by the Friedmann equation for the acceleration of the scale factor, and this for any inhomogeneous solutions. This means that there is no global backreaction of the inhomogeneities on the isotropic expansion.

If $\T\Xi \not=0$ and $\T\rot \not= 0$, the backreaction is
\begin{equation}
\CQ_\Sigma = -\Saverage{\Xi_{cd}\Xi^{cd}} + \Saverage{\omega_{cd}\omega^{cd}}. \label{eq::NEN_back_1}
\end{equation}
The fields $\T\Xi$ and $\T\omega$ are free, and especially they are decoupled from the fluid spatial velocity $\T v$. When deriving the Newton-Cartan equations as a limit of general relativity, we showed in \cite{2021_Vigneron} that $\T\omega$ must vanish due to geometrical constraints. Therefore we also expect this tensor to vanish if this NEN theory can be retrieved with a non-relativistic limit. We are let with the traceless-transverse tensor $\T\Xi$ which describes an anisotropic expansion and which is not constrained by the system~\eqref{eq::NEN_Conservation}-\eqref{eq::NEN_Kunzle}. So unless there is a hidden condition relating this tensor to $\T v$, coming from general relativity, it is not clear to us if we can consider that equation~\eqref{eq::NEN_back_1} corresponds to a physical backreaction.

If this NEN theory is the right one, then a Euclidean, spherical or hyperbolic closed universe, whose dynamics is locally Newtonian and the global expansion isotropic, has no global backreaction. For such a universe, no effects of the inhomogeneities can be observed on the evolution of the scale factor.\saut

\remark{In equation~\eqref{eq::NEN_back_1}, the backreaction $\CQ_\Sigma$ is defined as the deviation with respect to the Friedmann equation: i.e. $\CQ_\Sigma \coloneqq 3\left(\dot H + H^2\right) + 4\pi G \Saverage{\rho} - \Lambda$. Because the Raychaudhuri equation~\eqref{eq::NEN3_Einstein_ab} features an additional term, it is not algebraically equivalent to the Raychaudhuri equation of Newton's theory, and thus the usual formula ``$\CQ_\CD \coloneqq \frac{2}{3}\left(\Daverage{\theta^2} - \Daverage{\theta}^2\right) - \Daverage{\sigma_{cd}\sigma^{cd}} + \Daverage{\Omega_{cd}\Omega^{cd}}$'' is not valid anymore. Instead the definition $\CQ_\Sigma \coloneqq 3\left(\dot H + H^2\right) + 4\pi G \Saverage{\rho} - \Lambda$ leads to
\begin{equation*}
	\CQ_\Sigma = \frac{2}{3}\left(\Saverage{\theta^2} - \Saverage{\theta}^2\right) - \Saverage{\sigma_{cd}\sigma^{cd}} + \Saverage{\Omega_{cd}\Omega^{cd}} - \frac{\Rt}{3}\Saverage{v^cv_c}.
\end{equation*}}

\subsubsection{Space--time separation of the spatial metric.}
\label{sec::chatte}

Taking equation~\eqref{eq::NEN_Theta_hb} in the class of coordinates where $\T\beta = -\T v$ (Galilean coordinates), it becomes
\begin{equation}
	\derivtN{-\T v} h_{ab} + 2H(t)h_{ab} = \Xi_{ab}, \label{eq::NEN_sepa_met_1}
\end{equation}
which, in the case $\T\Xi = 0$, leads to
\begin{equation}
	h_{ab}(t,x^i) = a^2(t) \bar{h}_{ab}(x^i), \label{eq::NEN_sepa_met_2}
\end{equation}
where the Ricci curvature tensor $\bar{\T{\Rt}}$ related to the metric $\bar{\T h}$ is $\displaystyle \bar{\T{\Rt}} = \frac{\Rt_{\rm \bf i}}{3}\bar{\T h}$, with $\Rt_{\rm \bf i}$ the initial scalar curvature of $\T h$.

Equation~\eqref{eq::NEN_sepa_met_2} shows that the spatial metric components can be separated into space and time dependence, which was also the case in Newton's theory. This is a major property as it ensures that in Galilean coordinates, we can solve the system~\eqref{eq::NEN_sys_Ri}--\eqref{eq::NEN_sys_av_rho} assuming a time-independent background metric $\bar{\T h}$.

In the case $\T \Xi \not= 0$, the equation~\eqref{eq::NEN_sepa_met_1} does not lead to the separation of the metric components. From the uniqueness of the SVT decomposition, because $\T\Xi$ is a symmetric traceless-transverse tensor, it cannot be written as the symmetric gradient of a vector. This means that there is no coordinate system in which equation~\eqref{eq::NEN_Theta_hb} becomes $\derivtN{\T \beta} h_{ab} + 2H(t)h_{ab} = 0$. Therefore, the space and time separation of $\T h$ is only possible if $\T\Xi = 0$.

\subsubsection{$N$-body description.}

A major strength of Newton's theory is the fact that its system of equations can be written either as a partial differential system (fluid description), or as an ordinary differential system for the evolution equations (particle description). This results from the linearity of the Poisson equation. In the present proposed NEN theory, the same feature is present: the Green function of the Poisson equation~\eqref{eq::NEN_New_Gaus_1} is sufficient to describe the gravitational field of any distribution of matter. Therefore, we have an \textit{exact} $N$-body description:
\begin{enumerate}
	\item the gravitational field $\T g_{\rm tot}$, i.e. the solution of equation~\eqref{eq::NEN_New_Gaus_1}, for $N$ mass points described by Dirac fields $\{M_i\delta(\T x - \T x_i)\}_{i \in \llbracket 1,N\rrbracket}$, with $M_i$ and $\T x_i$ the mass and position of the $i^{\rm th}$ particle, is
\begin{equation}
	\T g_{\rm tot}(\T x) = \sum_i^N \T g_\Sigma(\T x - \T x_i),
\end{equation}
where $\T g_\Sigma$ is the solution of~\eqref{eq::NEN_New_Gaus_1} for a single Dirac field in $\Sigma$. We calculate it in the case $\Sigma = \mS^3$ in the next section.
	\item The trajectory of each point mass is given (for isotropic expansion) by Newton's second law~\eqref{eq::NEN_def_g_1}:
\begin{equation}
	\ddot{x_i}^a + 2H \dot{x_i}^a = g_{\rm tot}^a + (a_{\not= \rm grav})^a.
\end{equation}
\end{enumerate}
This is a major strength of the present NEN theory as $N$-body numerical simulations can be performed without any approximation with respect to the equations, something not possible with the first theory we proposed in section~\ref{sec::NEN_I} which \textit{a priori} cannot be reduced to a system of ordinary differential equations for the evolution equations.


\subsubsection{The mass point gravitational field.}
\label{sec::NEN_1bis_grav}

The $N$-body description requires the mass point solution of the Poisson equation~\eqref{eq::NEN_New_Gaus_1} which corresponds to the solution of
\begin{equation}
	\D_cg^c = -4\pi G\left[M\delta_\Sigma - \frac{M}{V_\Sigma}\right],
\end{equation}
where $\delta_\Sigma$ is the Dirac field of the manifold $\Sigma$.
For simplicity, we consider only the case where $\Sigma = \mS^3$, i.e. the 3-sphere. Then, choosing the hyperspherical coordinates~\eqref{eq::NEN_hyper_coord}, and the particle placed at the origin, we obtain
\begin{align}
	g^\xi = -\frac{G M}{\Rc^3\sin^2\xi} \left[1 - \frac{\xi}{\pi} + \frac{\cos\xi\sin\xi}{\pi}\right], \label{eq::NEN_grav_1bis}
\end{align}
where $\Rc(t) = a(t) R_{c, \rm \bf i} = \sqrt{6/\Rt}$ is the curvature radius of the 3-sphere and $R_{c, \rm \bf i}$ its initial value. Note that because we do not use normalised coordinate basis vectors, we have $\T g \cdot \T g = \Rc^2(g^\xi)^2$, which means that $\Rc g^\xi$ corresponds to the measured gravitational field strength.

The behaviour of the gravitational field at the poles is ($\xi = 0$ is the particle position and $\xi = \pi$ is the opposite pole)
\begin{align}
	\Rc g^\xi(\xi \sim 0)		&= -\frac{GM}{r_0^2} - \frac{GM}{3\Rc^2} + \mathcal{O}(r_0), \label{eq::NEN_g_lim} \\
	\Rc g^\xi(\xi \sim \pi)	&= -\frac{2GM}{3\pi\Rc^3}r_\pi + \mathcal{O}\left(r_\pi^3\right), \label{eq::NEN_g_lim_pi}
\end{align}
where $r_0 := \Rc \xi$ is the distance to the origin and $r_\pi := \Rc (\pi -\xi)$ the distance to the opposite pole. The first problem of the NEN theory proposed by \cite{2009_Roukema_et_al,2020_Barrow} is solved: the field at the opposite pole $\xi = \pi$ of a point mass particle does not diverge anymore. It is a first order repulsive field, and therefore tends to zero when $\xi \rightarrow \pi$. This is physically coherent.

The second term in the right-hand-side of equation~\eqref{eq::NEN_g_lim} is an additional term with respect to Newton's gravitational law. It is called \textit{topological acceleration}. Such a zeroth order term is also present in Newton's theory (but in a different form) in the case $\Sigma$ is a 3-torus \cite{2007_Roukema_et_al}, instead of $\mathbb{R}^3$ with fall-off conditions at infinity. Therefore it is not necessarily linked to non-Euclidean topologies, but also to non-$\mathbb{R}^3$ topologies.\saut


\subsubsection{Discussion.}
\label{sec::bite}

This NEN theory has none of the limitations we listed for the first one in section~\ref{sec::NEN1_limitations}: the spatial metric is separable in space and time (if $\T\Xi = 0$ which is the case if the expansion is isotropic); no gravitomagnetic field is present (if $\T\rot = 0$ which is expected from general relativity); $N$-body description is possible. Furthermore, it solves the first problem of the NEN theory proposed by \cite{2009_Roukema_et_al,2020_Barrow}.

However, the curvature term in the modified Newton-Cartan equations still requires the presence of a fluid to be defined. This contrasts with the gravitational system~\eqref{eq::NEN_def_g_1}-\eqref{eq::NEN_sys_av_rho} resulting from these equations, which is defined even in the absence of a fluid. This suggests that we could define a Galilean observer independently of a fluid. We propose the following covariant definition:\saut

\begin{mdframed}
\definition{A \textbf{Galilean observer} (in a 4-manifold equipped with a Galilean structure) is define by a timelike vector $\T G$ constrained by the following equations:
\begin{align}
\begin{cases}
	h^{\mu[\alpha} \nabla_\mu G^{\beta]} = 0, \\
	\nabla_\mu\left( h^{\nu(\mu} \nabla_\nu G^{\alpha)} - \frac{1}{3}\nabla_\nu G^\nu h^{\mu\alpha}\right) = 0, \\
	h^{\alpha\mu}R_{\mu\nu}G^\nu = 0.
	\end{cases}\label{eq::def_G}
\end{align}}
\end{mdframed}
The second condition is equivalent to saying that the shear tensor $\TT{\sigma}{G}^{ab} \coloneqq \TT{\Theta}{G}^{ab} - \frac{1}{3}\TT{\theta}{G} h^{ab}$ of the Galilean observer is transverse, i.e. $D_c\TT{\sigma}{G}^{ca} = 0$, where $\TT{\Theta}{G}^{ab}$ is the expansion tensor of $\T G$, and $\TT{\theta}{G}$ its trace, defined equivalently as for $\T u$ in section~\eqref{sec::caca}. This condition implies that in a SVT decomposition of $\TT{\Theta}{G}_{ab}$, the term $\D_{(a}v_{b)}$ (see section~\eqref{eq::NC_decomp}) is zero.

\eqref{eq::def_G} is an explicitly covariant definition of a Galilean observer. It is equivalent to the following two conditions: (i) in coordinates adapted to the foliation defined by $\T\tau$ and such that $\T\partial_t = \T G$, the time derivative of the spatial components of the space metric have the form
\begin{equation}
	\frac{1}{2}\partial_t h^{ij} = -H(t) h^{ij} - \Xi^{ij}, \label{eq::h_ij_Gal}
\end{equation}
with $\Xi^{ij}$ being traceless-transverse and (ii) $\T G$ is vorticity free with respect to the Galilean structure, i.e.
\begin{equation}
	h^{\mu[\alpha}\nabla_\mu G^{\beta]} = 0. \label{eq::Omega_Gal}
\end{equation}

This definition is close to the one of ``Newtonian observers'' in \cite{1998_vanElst_et_al}, where the translation of their definition in the framework of Galilean manifolds would give ``$h^{\mu[\alpha} \nabla_\mu G^{\beta]} = 0 = h^{\mu(\alpha} \nabla_\mu G^{\beta)} - h^{\alpha\beta}\nabla_\mu G^\mu/3$''. In other words, the last conditions in~\eqref{eq::def_G} are replaced by a shear free condition. This definition is a bite stronger than ours as it imposes $\T\Xi = 0$ implying that expansion in a Newtonian-like theory should be isotropic. Therefore, in absence of a condition obtained from a limit of general relativity that would impose $\T\Xi = 0$, formula~\eqref{eq::def_G} is the definition of a Galilean observer to use.


As for now we have not found a more compact equation which would be equivalent to the three conditions~\eqref{eq::def_G}. With this definition, we retrieve the same gravitational system~\eqref{eq::NEN_def_g_1}-\eqref{eq::NEN_sys_av_rho} but the spacetime equations of the theory, and in particular the term $\frac{\Rt}{3}\bb{G}_{\alpha\beta}$, are well defined in vacuum with $\T u = 0$. The only requirement of the theory is the existence of such a Galilean observer, which must be justified from general relativity or be an additional physical property in the definition of Galilean structures.\\

\remark{In order for $\Xi_{ij}$ to describe only anisotropic expansion, we expect this tensor to be gradient free, i.e. $D_k\Xi_{ij} = 0$. Having just a divergence free condition would allow $\T\Xi$ to have local dynamics. Therefore, a better definition for a Galilean observer might be to impose the full expansion tensor of $\T G$ to have a zero spatial gradient: $h^{\gamma\mu}\nabla_\mu \left[h^{\nu(\alpha} \nabla_\nu G^{\beta)}\right] = 0$.}

\section{Conclusion}

In this paper we presented two novel approaches to define a non-Euclidean Newtonian theory in either spherical or hyperbolic classes of topologies. These approaches consider the Galilean invariance as a fundamental principle of a Newtonian-like theory, regardless of the topology. They are based on a minimal modification of the Newton-Cartan equation allowing for a non-Euclidean topology. We derived the expansion law in each case, thus generalising the \BET. The main properties of these two theories are summarised in table~\ref{tab:NEN}.

In the first approach, a backreaction appears which depends, in particular, on the averaged specific energy of the fluid (an expected behaviour of backreaction), and also on the type of topology. However because the spatial metric is non-separable, no $N$-body calculation is possible and a gravitomagnetic field is present, this theory is far from what we \textit{a priori} expect of a Newtonian-like theory.

The second approach has the same expansion law as in Newton's theory. Then in the assumption of an isotropic global expansion, there is no cosmological backreaction. This extends the conclusions of the {\BET} to spherical and hyperbolic topologies. In this theory, $N$-body calculation can be performed, along with the possibility of separating the spatial metric in space and time, i.e. having a background description.

Without a justification from a non-relativistic limit of general relativity, the question of the ``right'' non-Euclidean Newtonian theory remains open, as well as the generalisation of the Buchert-Ehlers theorem. Because of the many similarities of the second proposed theory in section~\ref{sec::NEN_NC_app} with the (Euclidean) Newtonian theory, we expect it to be the ``right'' one, and its equations [equations~\eqref{eq::NEN_def_g_1}-\eqref{eq::NEN_sys_av_rho}] should be used to perform $N$-body simulations of structure formation in a universe with a non-Euclidean topology. For this purpose, in \citep{2022_Vigneron_et_al} we calculate the gravitational potential in different spherical topologies using this NEN theory.

Follow-up work is dedicated to further study the second NEN theory and to develop a limit of general relativity allowing us to derive the ``right'' NEN theory \citep{2022_Vigneron_c}.

\section*{Acknowledgements}

This work has received funding from the Center of Excellence of Nicolaus Copernicus University in Toru\`n. I would like to thank Boudewijn Roukema for making me discover the world of non-Euclidean Newtonian theories during my master thesis in Toru\'n. I also thank the referees for their very constructive remarks.

\newpage

\begin{landscape}
\hspace{2cm}
\begin{table}
\centering
\vspace{3cm}
\begin{tabular}{m{5.2cm}m{6.8cm}m{2.4cm}m{1.9cm}m{5cm}}
\hline
\centering\bf NEN theory & \centering\bf $3\left(\dot H + H^2\right) + 4\pi G \Saverage{\rho} - \Lambda =$ & \centering\bf $N$-body & \centering\bf Separable spatial metric & \centering\arraybackslash\bf Gravitomagnetism \\
\hline\hline
Based on the NC equation: \begin{equation*}\Rf_{ab} - \bb{u}_{ab} \frac{\Rt}{3} = 4\pi G\rho\, \tau_a\tau_b\end{equation*} (Fluid observer term) & \vspace{-1cm}\centering
		\[- \Saverage{\Xi_{cd}\Xi^{cd}} + \Saverage{\omega_{cd}\omega^{cd}}\]
		\[\displaystyle - \frac{\Rt}{3}\Saverage{v_cv^c} + \frac{\Rt}{6}\left(\Saverage{\Phi_v^2} - \Saverage{\Phi_v}^2\right)\]
		\[+ \Saverage{\tensor[^\Delta]{\Omega}{_c_d}\tensor[^\Delta]{\Omega}{^c^d}}\]
	with $v_a = \D_a\Phi_v + \epsilon_{acd}\D^cA^d$
& \centering No & \centering No, unless $\T\Xi = 0$ \underline{and} $\Phi_v = 0$ &  \centering\arraybackslash Yes: $\GM_{ab} \coloneqq \tensor[^\Delta]{\Omega}{_a_b} + \omega_{ab}$
		\[\D_{[a}g_{b]} = -\partil_t \GM_{ab}\]
		\[\D^c \GM_{ca} = J_a\]
	with $J_a \coloneqq \frac{\Rt}{3} \epsilon_{acd}\D^cA^d$ and no displacement current  \\
\hline
Based on the NC equation: \begin{equation*}\Rf_{ab} - \bb{G}_{ab} \frac{\Rt}{3} = 4\pi G\rho\, \tau_a\tau_b\end{equation*} (Galilean observer term) & \centering $- \Saverage{\Xi_{cd}\Xi^{cd}} + \Saverage{\omega_{cd}\omega^{cd}}$ & \centering Yes, gravitational field given by~\eqref{eq::NEN_grav_1bis} in $\mS^3$ & \centering Yes, if $\T\Xi = 0$ & \centering\arraybackslash No, if $\T\omega = 0$ \\
\hline

\end{tabular}
\caption{Summary of the proposed NEN theories. The second column quantifies the difference with an FLRW expansion. For the sake of generality we kept the fields $\T\Xi$ and $\T\rot$ non zero. However, as in Newton's theory (see appendix B in \cite{2021_Vigneron}) we expect $\T\omega = 0$ if the NEN theory is derived from a non-relativistic Galilean limit. For an isotropic expansion, we also expect $\T\Xi = 0$. In that case, only the first NEN theory features un non-zero cosmological backreaction and a departure from the $\Lambda$CDM scenario.}
\label{tab:NEN}
\end{table}
\end{landscape}

\section*{References}
 {When available the bibliography style features three different links associated with three different colors: links to the journal/editor website or to a numerical version of the paper are in \textcolor{LinkJournal}{red}, links to the ADS website are in \textcolor{LinkADS}{blue} and links to the arXiv website are in \textcolor{LinkArXiv}{green}.}
 
\bibliographystyle{QV_mnras}
\IfFileExists{bib_General.bib}{\bibliography{bib_General}}{\bibliography{/Users/quentinvigneron/Documents/Travail/Research/tex_/bib_General}}

\appendix

\end{document}